\documentclass[trackchanges]{aastex701}

\usepackage{multirow}
\usepackage{booktabs}
\usepackage{subcaption}

\begin{document}

\title{Plate-YOLO: An End-to-End Object Detection Framework for Robust Astrometric Registration of Historical Plates}

\correspondingauthor{Shiyin Shen, Yong Yu}
\author[orcid=0000-0002-9572-6212]{Quanfeng Xu}
\affiliation{Shanghai Astronomical Observatory, Chinese Academy of Sciences, 80 Nandan Rd., Shanghai 200030, PR China}
\affiliation{School of Astronomy and Space Science, University of Chinese Academy of Sciences, 1 East Yanqi Lake Rd., Beijing 100049, PR China}
\email[show]{xuquanfeng@shao.ac.cn}  

\author{Zhengjun Shang}
\affiliation{Shanghai Astronomical Observatory, Chinese Academy of Sciences, 80 Nandan Rd., Shanghai 200030, PR China}
\email{zjshang@shao.ac.cn}  

\author[orcid=0000-0002-3073-5871]{Shiyin Shen}
\affiliation{Shanghai Astronomical Observatory, Chinese Academy of Sciences, 80 Nandan Rd., Shanghai 200030, PR China}
\affiliation{Key Lab for Astrophysics, Shanghai, 200234, China}
\email[show]{ssy@shao.ac.cn}

\author{Yong Yu}
\affiliation{Shanghai Astronomical Observatory, Chinese Academy of Sciences, 80 Nandan Rd., Shanghai 200030, PR China}
\affiliation{School of Astronomy and Space Science, University of Chinese Academy of Sciences, 1 East Yanqi Lake Rd., Beijing 100049, PR China}
\email[show]{yuy@shao.ac.cn}

\author{Mingyang Ma}
\affiliation{Institute for Frontiers in Astronomy and Astrophysics, Beijing Normal University, Beijing, 102206, PR China}
\affiliation{School of Physics and Astronomy, Beijing Normal University, Beijing, 100875, PR China}
\email{mamingyang@mail.bnu.edu.cn}  

\author{Zhenghong Tang}
\affiliation{Shanghai Astronomical Observatory, Chinese Academy of Sciences, 80 Nandan Rd., Shanghai 200030, PR China}
\affiliation{School of Astronomy and Space Science, University of Chinese Academy of Sciences, 1 East Yanqi Lake Rd., Beijing 100049, PR China}
\email{zhtang@shao.ac.cn}  

\author{Meiting Yang}
\affiliation{Shanghai Astronomical Observatory, Chinese Academy of Sciences, 80 Nandan Rd., Shanghai 200030, PR China}
\email{ymt@shao.ac.cn}  

\author{Jing Yang}
\affiliation{Shanghai Astronomical Observatory, Chinese Academy of Sciences, 80 Nandan Rd., Shanghai 200030, PR China}
\email{jyang@shao.ac.cn}  

\author{Jianhai Zhao}
\affiliation{Shanghai Astronomical Observatory, Chinese Academy of Sciences, 80 Nandan Rd., Shanghai 200030, PR China}
\email{zhaojh@shao.ac.cn}  

\begin{abstract}

Historical astronomical plates provide a long temporal baseline for time-domain astrophysics, but degradation and structural artifacts can hinder automated astrometric calibration. We present Plate-YOLO, a framework that combines source detection and stellar classification in one detector and supplies source catalogs to a separate astrometric solver. The detector adapts YOLOv5 with Space-to-Depth feature rearrangement, SimAM attention, a high-resolution P2 head, and a Normalized Wasserstein Distance loss for small bounding boxes. A Gaia-guided fine-tuning procedure uses initially solved plates to construct labels for a new archive. On 150 UBAI plates for which the conventional pipeline had failed, the Chinese-trained model solved 40 plates, and target-archive fine-tuning recovered a further 106, giving an overall recovery rate of 146/150 (97.3\%). In the reported hardware configuration, source extraction took 6.61 s per plate for Plate-YOLO and 16.73 s for \textsc{SExtractor}. These results demonstrate the practical value of adapting source detection to degraded historical plates.

\end{abstract}



\section{Introduction} 

Historical astronomical photographic plates serve as a unique and irreplaceable record of the dynamic universe, providing a century-long baseline essential for time-domain studies and understanding long-term astronomical phenomena. For instance, the Harvard DASCH project \cite[][]{Grindlay_Tang_Los_Servillat_2011} utilized custom high-speed scanners \cite[][]{simcoe2006ultrahigh} to digitize approximately 450,000 plates. Similarly, the APPLAUSE project \cite[][]{2024A&A...687A.165E} integrated over 70,000 historical plates from multiple German observatories. On a larger volumetric scale, the SuperCOSMOS project \cite[][]{10.1111/j.1365-2966.2001.04660.x, 10.1111/j.1365-2966.2001.04661.x} digitized southern-hemisphere Schmidt plates to produce sub-arcsecond precision reference catalogs. Meanwhile, specialized efforts like the Paris Observatory's PADC project \cite[][]{Robert_2021} focus on high-precision digitization tailored for Solar System kinematics. In parallel with these international endeavors, a comprehensive effort in China has centralized and digitized approximately 30,000 historical plates at the Shanghai Astronomical Observatory using custom high-precision scanners \cite[][]{yangmeiting, Yu_2017}.

\begin{figure}
    \includegraphics[width=\linewidth]{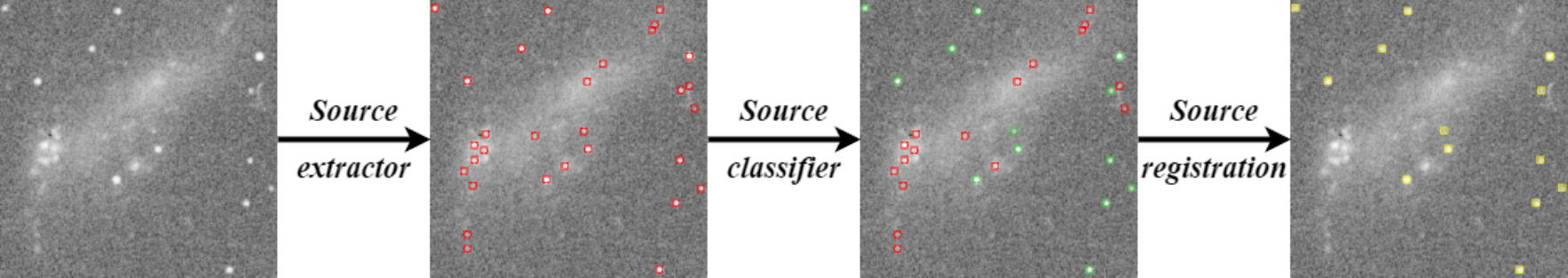}
    \caption{
    An overview of the astrometric registration workflow for historical nighttime astronomical plates \cite[Example for Chinese plate BJ8703SD2899001, adapted form][]{Xu_2026}. 
    }
    \label{fig:traditional_method}
\end{figure}

To fully realize the scientific goals of these digitized plate archives, the fundamental step is accurate astrometric calibration, especially for those plates lacking reliable pointing information. As illustrated in Fig.~\ref{fig:traditional_method}, the conventional calibration pipeline relies on a sequential paradigm: full-field source detection across the large image, followed by detailed classification and astrometric registration of stellar objects from extracted cutouts. The initial source detection stage  typically relies on threshold-based algorithms like \textsc{SExtractor} \citep{Bertin1996}, e.g. \cite[][]{Shang_2024} and \cite[UBAI,][]{10.1093/mnras/stag335}. Once validated stellar sources are extracted, they can be fed into blind solvers like \texttt{Astrometry.net} \citep{Lang_2010} for a coarse match, followed by a fine World Coordinate System (WCS) solution via cross-matching with modern high-precision catalogs like Gaia.

However, long-term preservation challenges frequently result in plates that are substantially degraded by complex structural noise and a variety of artifacts, including scratches, mold growth, and emulsion defects. To reduce the influence of these degradations on subsequent analysis, researchers have introduced additional stellar source classification procedures before the astrometric registration stage. For example, \cite{Shang_2024} utilized a Support Vector Machine (SVM) to classify and filter sources based on the measurement parameters extracted by \textsc{SExtractor}. Taking this a step further, \cite{Xu_2026} avoided reliance on hand-crafted source parameters by applying a Swin Transformer deep learning classifier directly to the image cutouts generated by \textsc{SExtractor}, demonstrating a significant improvement over traditional classification methods in filtering out artifacts.

Despite the success of such advanced classifiers, initial source detection in these pipelines still relies on traditional methods that require manual parameter tuning \citep{Lang_2010,Masias_2012}. As modern astronomy enters an era of massive data from next-generation surveys, automated deep learning techniques \citep{Ntampaka2021,Fluke2020,VillaescusaNavarro2021} offer a more streamlined solution. For example, object detection architectures such as Faster R-CNN \citep{Ren2015}, SSD \citep{Liu2016}, and YOLO (You Only Look Once) \citep{Redmon2016,Redmon2018} are capable of performing source detection and classification simultaneously in a single step. These frameworks have also been successfully applied in a variety of astronomical contexts. For optical imaging, \cite{Grishin2023} successfully introduced YOLO-CL to identify galaxy clusters in large-scale optical surveys. Similarly, AstroYOLO \citep{He2023} integrated a hybrid CNN-Transformer architecture to capture global feature dependencies for detecting blue horizontal-branch stars in photometric images. STAR-YOLO \citep{Zhu_2025} employed multi-scale attention mechanisms and an Adaptive Focal Spatial-IoU loss function to pinpoint irregular and small-target galaxies accurately. Beyond optical data, these YOLO frameworks also perform well in the radio domain \citep{Cornu2025}. These studies collectively demonstrate that end-to-end deep learning effectively bypasses multi-step procedural dependencies, offering a highly robust approach for source identification and classification under complex observational conditions.

Historical plates pose additional challenges for learned object detection because their image quality, instrumental response, and artifacts vary substantially across archives. Reliable annotations are difficult to obtain for degraded images. Differences in stellar image profiles, photographic emulsions, spatial resolution, and preservation conditions can also reduce performance when a detector is transferred between telescopes or archives \citep{Xu_2026}. Plate-YOLO addresses these challenges through small-object detection and archive-specific fine-tuning.

In this paper, we propose Plate-YOLO, an object detection framework designed to facilitate the robust astrometric registration of historical photographic plates. Plate-YOLO streamlines the conventional pipeline by unifying source extraction and source classification into a single stellar object detection step. We initially train the model on well-characterized Chinese historical plates to establish a reliable baseline \cite[][]{Shang_2024, Xu_2026}, and then apply it directly to the  Ulugh Beg Astronomical Institute (UBAI) archive to evaluate its cross-archive generalization. To account for the inherent physical and observational differences between these archives, we introduce a Gaia-guided fine-tuning strategy that utilizes initially calibrated UBAI plates to generate accurate local reference labels. Ultimately, this pipeline enables successful stellar extraction and astrometric calibration for severely degraded plates that conventional methods fail to process.

This paper is organized as follows: Section \ref{sec:data} describes the Chinese and UBAI plate datasets and the preprocessing procedure. Section \ref{sec:method} introduces the Plate-YOLO framework, including the network architecture, the Gaia-guided fine-tuning strategy, and the photometric estimation. To systematically validate the specific architectural modifications of Plate-YOLO, a detailed ablation study is provided in Appendix \ref{sec:ablation}.
Section \ref{sec:experiments} establishes the baseline detection capabilities on the Chinese archive, and demonstrates the model's cross-domain generalization on the UBAI collection, highlighting the successful astrometric recovery of previously failed plates. Section \ref{sec:discussion} discusses the practical advantages and current limitations of our approach. Finally, Section \ref{sec:conclusion} summarizes the core conclusions.

\section{Observational Data and Pre-processing}\label{sec:data}

To validate the effectiveness of the proposed deep learning framework, we utilize highly heterogeneous astronomical plates from two distinct historical archives. We also describe the pre-processing pipeline to construct standardized training samples from these raw digitized images (Section \ref{sec:proprocess}).

\subsection{The Chinese Astronomical Plates}\label{sec:China}

The training data come from China's digitized historical night-time astronomical plates \cite[][]{Shang_2024}. These observations span more than one century and were obtained with 11 telescopes at five stations, with apertures from 15 cm to 1.56 m \citep[see details in][]{Xu_2026}.

The training sample comprises 2,000 full plates that had previously been astrometrically calibrated using the conventional pipeline. These solutions support the generation of Gaia-matched reference labels, whose limitations are discussed in Section~\ref{sec:cutout}. Furthermore, we randomly selected 1,000 similarly calibrated plates from the remaining archive: 500 plates formed a validation set to monitor model convergence, and 500 plates constituted a test set for final performance evaluation.

\subsection{The UBAI Astronomical Plates}

To evaluate the practical application of our model, we utilized independent observational data from the UBAI historical plate archive in Uzbekistan. These observations provide a long-term observational record spanning from 1895 to 2005. They were predominantly collected using two distinct wide-field optical systems, the Tashkent Normal Astrograph (TNA) and the Double Astrograph Zeiss (DAZ), with apertures of 0.33 and 0.40 m respectively. The archive was recently digitized using SHAO digitizers \cite[][]{yangmeiting,Shang_2026} and comprises over 15,000 photographic plates, encompassing single-exposure, multiple-exposure, and grating fields. \cite{10.1093/mnras/stag335} analyzed 2,574 selected single-exposure plates from this archive employing standard astrometric reduction pipelines. Of these, 2,330 plates were successfully astrometrically calibrated, yielding mean positional uncertainties of 0.278$''$ for TNA and 0.287$''$ for DAZ\footnote{Of the 244 plates that initially failed conventional calibration, later pipeline optimizations recovered some \cite[private communication from][]{10.1093/mnras/stag335}, leaving 150 highly degraded plates that failed all conventional methods.}.

\subsection{Data Pre-processing and Label Generation}\label{sec:proprocess}

To convert raw digitized plates into a structured training dataset, our data preparation comprises two primary operations. Section \ref{sec:non_linear} describes the non-linear transformation used to compress the dynamic range of the original images. Section \ref{sec:cutout} outlines the spatial partitioning of the full-field plates into sub-image cutouts and the automated generation of their corresponding ground-truth labels.

\subsubsection{Non-linear Dynamic Range Compression}\label{sec:non_linear}

The digitized historical plates are stored as 16-bit FITS images. \texttt{The single-band astronomical data} pixel intensity distributions exhibit strong positive skewness, which reduces the effectiveness of conventional CNN-based methods when applied directly to the raw digitized data. To mitigate this issue, we employ a non-linear intensity compression procedure that remaps the pixel values into a dynamic range characteristic of natural images.

Pixel values in each single-band patch were clipped at the 1st and 99th percentiles, normalized to $[0,1]$, and stretched using an inverse hyperbolic sine transformation:
\begin{equation}
\label{equ:asinh}
I_{\rm out}=\frac{\mathrm{asinh}\left[\beta\frac{I_{\rm clip}-P_{1}}{P_{99}-P_{1}}\right]}{\mathrm{asinh}(\beta)},\quad \beta=10.
\end{equation}
where $I_{\rm clip}$ is the clipped pixel value, $P_{1}$ and $P_{99}$ are the corresponding percentiles computed independently for each patch, and $\beta=10$ controls the stretch.

The $\mathrm{asinh}$ mapping is well suited to the large dynamic range of modern astronomical imaging surveys \citep{Lupton_1999}: it behaves approximately linearly at low signal levels, thereby preserving the morphology of faint PSF profiles, while approaching a logarithmic response for bright or saturated regions, thus preventing their dominance in the dynamic range. For clarity, the non-linearly transformed image is used solely for source detection, while all brightness-proxy measurements use the original, unstretched FITS values.

\subsubsection{Spatial Windowing and Labels}\label{sec:cutout}

Digitized historical astronomical plates routinely exceed $8000 \times 8000$ pixels, making direct deep learning ingestion computationally intractable. To prepare the data for Plate-YOLO, we implemented an overlapping sliding-window strategy, partitioning the full-field images into $256 \times 256$ pixel sub-image tensors. A 26-pixel overlap (approximately 10\% of the window width) reduces truncation near window boundaries. After detection in each patch, local bounding boxes are mapped back to the global plate coordinate system of the original plate.

We applied this partition to construct a representative dataset $\mathcal{D}_{s} = \{(x_j^s, y_j^s)\}_{j=1}^{N_s}$, where $x^s$ represents the image tensor and $y^s$ encapsulates both the bounding box geometries and class labels. To generate $y^s$ objectively, we implemented an automated pipeline.

First, \textsc{SExtractor} was run with a detection threshold of $1.5\sigma$ above the local background to obtain candidate sources. This low threshold increases sensitivity but does not guarantee completeness. Training-box centers were taken from the windowed positions (\texttt{XWIN\_IMAGE}, \texttt{YWIN\_IMAGE}), and the box sizes were defined as $w=2\times\texttt{A\_IMAGE}$ (semi-major) and $h=2\times\texttt{B\_IMAGE}$ (semi-minor axes).

Candidate detections were cross-matched with Gaia DR3 \citep{2023A&A...674A...1G} using reference sources with $G\leq18$ mag \footnote{We found that entries with Gaia $G>18$ account for less than 0.00336\% of entries with valid $G$ magnitudes in the photometric catalogs examined \citep{Ma_2025}.}, whose coordinates were propagated to each plate epoch using Gaia proper motions. Detections with a counterpart within $3\times{\rm RMS}$ of the plate astrometric solution were labeled as stellar, whereas unmatched detections were operationally labeled as non-stellar. Positional errors, blending, and catalog selection can also affect matching. These labels therefore provide a catalog-based reference rather than a guarantee of photometric completeness.

As described in Section~\ref{sec:China}, the dataset was divided into independent training, validation, and test sets at the full-plate level (see Table~\ref{tab:data_split} in detail). For computationally efficient model development, we sampled 100 patches from each training plate, yielding 200,000 training cutouts containing 645,307 stellar and 250,259 fake source label instances. We similarly constructed the validation set using a fixed sample of 100 patches per plate to monitor model convergence. The test set was constructed by selecting 500 patches from each test plate. All label counts in Table~\ref{tab:data_split} refer to patch-level instances, including repeated appearances of the same source in overlapping patches. Detection metrics and ablation results are evaluated per patch, while the full-plate split prevents data leakage.

\begin{table}[h]
	\centering
	\caption{Dataset accounting for the Chinese archive.}
	\label{tab:data_split}
    \begin{tabular}{lrrrr}
    \hline
    Split & Full plates & Cutouts & Stellar & Fake source \\
    \hline
    Training & 2,000 & 200,000 & 645,307 & 250,259 \\
    Validation & 500 & 50,000 & 148,941 & 61,511 \\
    Test & 500 & 250,000 & 847,812 & 405,188 \\
    \hline
	\end{tabular}
\end{table}

For full-plate inference in the astrometric pipeline, each plate was tiled into $256\times256$-pixel patches with a stride of 230 pixels, with the final windows shifted to the right and bottom boundaries to ensure complete coverage without padding. Predictions below a confidence threshold of 0.5 were discarded, and the remaining bounding boxes were transformed to global plate coordinates. Detections were pooled and deduplicated using class-wise global non-maximum suppression with an intersection-over-union threshold of 0.2. The centers and instrumental-magnitude rankings of the retained stellar detections were then passed to \texttt{Astrometry.net} as the source catalog. This global merging and deduplication procedure was used to construct full-plate source catalogs and was not applied to the patch detection evaluation.

In the Chinese training set, 37.2\% of stellar-class bounding boxes have both widths and heights no greater than 8 pixels, increasing to 61.5\% at a threshold of 12 pixels. On a stride-8 feature map, these thresholds correspond to just 1 and 1.5 grid spacings, respectively, limiting the spatial detail available for feature extraction and localization. This prevalence of small sources motivates the use of higher-resolution features and a regression loss tailored to small bounding boxes.

\section{Plate-YOLO Framework}\label{sec:method}

This section introduces the \texttt{Plate-YOLO} framework. Unlike conventional methods that rely on a disjointed two-step process (initial source extraction followed by subsequent stellar classification), Plate-YOLO unifies these tasks into a single, end-to-end object detection step. First, we detail the network's structural design and specialized loss functions. Next, a cross-archive refinement strategy is presented to ensure the detector's robust generalization across heterogeneous historical plate collections. Finally, we introduce an integrated aperture-based brightness estimation, which generates the relative brightness rankings required to seamlessly interface our deep learning extractions with classical astrometric solvers.

\subsection{Plate-YOLO Architecture}

\begin{figure*}
    \centering
    \includegraphics[width=\textwidth]{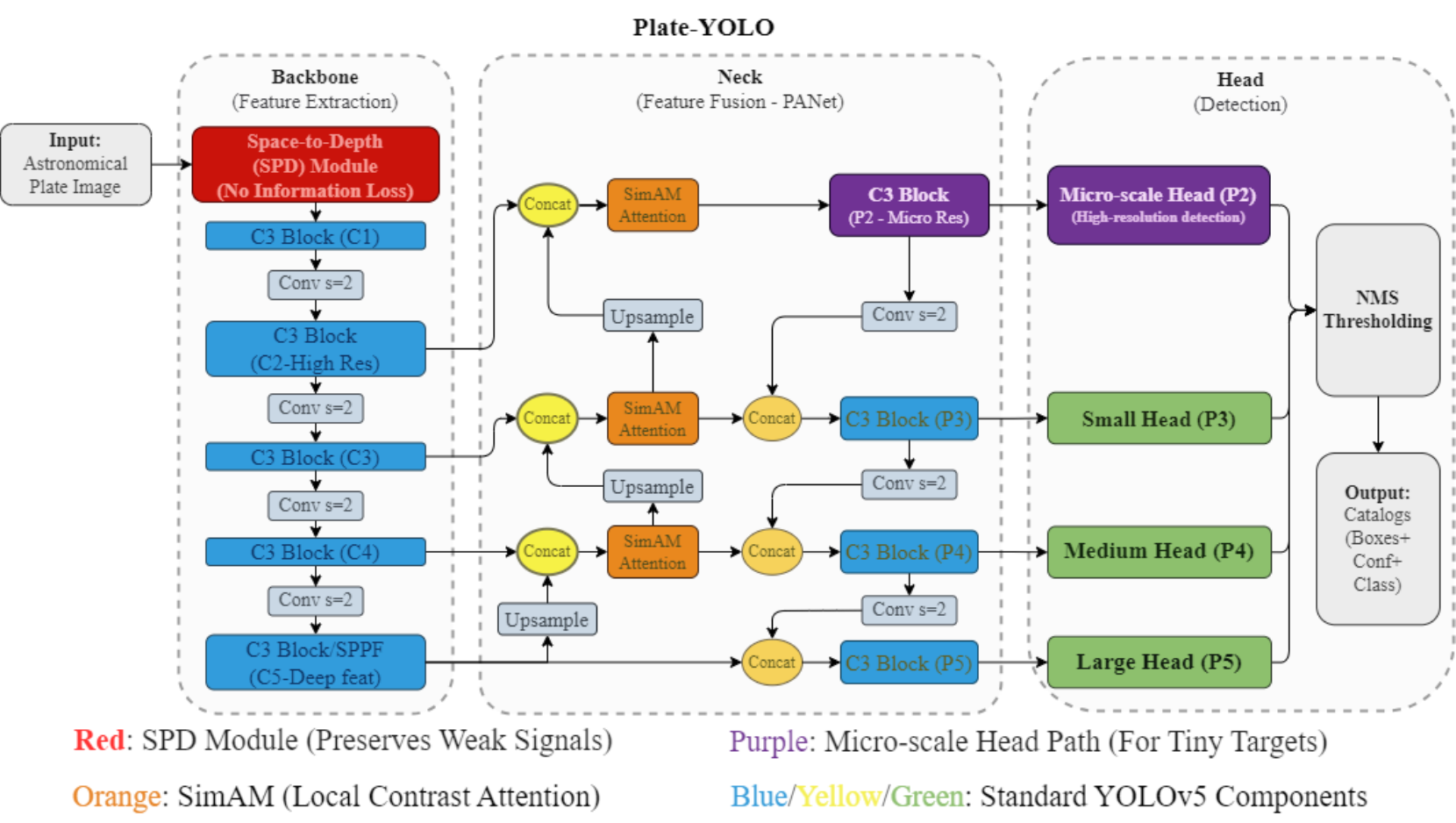}
    \caption{An overview of the Plate-YOLO architecture, highlighting the integration of the Space-to-Depth (SPD) module, SimAM attention mechanisms, and the dedicated Micro-scale Head (P2).}
    \label{fig:plate_yolo}
\end{figure*}

Plate-YOLO adapts the YOLOv5 baseline \cite[][]{khanam2024yolov5deeplookinternal} with three structural modifications tailored for micro-scale targets (Figure \ref{fig:plate_yolo}). To finalize predictions, we apply a confidence threshold to reject background noise, followed by Non-Maximum Suppression (NMS) to resolve overlapping bounding boxes. The effectiveness of these modifications is validated in Appendix \ref{sec:ablation}.

\subsubsection{Space-to-Depth module}\label{sec:spd}

In standard convolutional networks, stride-2 max pooling or strided convolutions are commonly used to reduce spatial dimensions and increase the receptive field. \cite{10.1007/978-3-031-26409-2_27} noted that this process can remove fine-grained spatial cues that are critical for detecting very small objects. For celestial targets spanning only a few pixels, this information loss is highly detrimental.

To mitigate this problem, Plate-YOLO modifies the early feature extraction hierarchy by introducing a Space-to-Depth (SPD) module \cite[][]{10.1007/978-3-031-26409-2_27} into the shallow layers of the CSPNet backbone \cite[][]{wang2020cspnet}. Unlike stride-2 convolution or max pooling, which reduce resolution by sampling or aggregating local neighborhoods, SPD performs a deterministic rearrangement of spatial information. Given an input feature map of size $(H, W, C)$ (e.g., $C=1$ for standard grayscale astronomical plates), the SPD module partitions each $2 \times 2$ spatial neighborhood into four sub-features and concatenates them along the channel dimension, producing an output of size $(H/2, W/2, 4C)$. This operation reduces the spatial resolution without discarding the original pixel-level responses. As a result, fine-grained cues from micro-scale celestial targets are preserved in the channel representation, providing richer shallow features for subsequent multi-scale fusion.

\subsubsection{Simple Parameter-Free Attention Module}\label{sec:simam}

While the SPD module preserves the spatial topology of faint signals, these features must still be disentangled from the complex structural background. Standard Convolutional Neural Networks (CNNs), constrained by their local receptive fields, can struggle to differentiate genuine point sources from similarly sized artifacts.  To suppress these structural irregularities and amplify genuine stellar signals, we embed a Simple Parameter-Free Attention Module \cite[SimAM,][]{Yang2021SimAMAS} at the multi-scale feature fusion nodes within the PANet-based neck network. Unlike conventional attention mechanisms \cite[e.g., Squeeze-and-Excitation,][]{8578843} that require fully connected layers and thousands of additional learned parameters, SimAM calculates 3D attention weights by evaluating an energy-based measure of the separability of each activation from other activations in the same feature channel.

SimAM reweights feature activations using channel-wise spatial statistics \citep{Yang2021SimAMAS}. This may help the detector distinguish sources from structured backgrounds. Its effect is evaluated through the detection metrics in Appendix~\ref{sec:ablation}.

\subsubsection{Micro-Scale High-Resolution Detection Head}\label{sec:p2}

The standard YOLOv5 architecture uses a three-tier detection head designed to capture large, medium, and small objects corresponding to the levels of the feature map $P5$, $P4$, and $P3$. However, astronomical targets on historical plates often fall significantly below the threshold of what computer vision algorithms classify as "small" objects.

Plate-YOLO adds a high-resolution P2 detection head. It combines shallow features with upsampled deeper features to provide a finer spatial grid for localization. This design is intended to improve sensitivity to sources occupying only a few pixels.

\subsection{Joint Loss Function}

The Plate-YOLO training objective combines bounding-box regression, objectness (foreground vs. background), and class prediction (stellar vs. fake sources):

\begin{equation}\label{equ:loss}
\mathcal{L}_{total} = \lambda_{box\_NWD}\mathcal{L}_{box\_NWD} + \lambda_{obj}\mathcal{L}_{obj} + \lambda_{cls}\mathcal{L}_{cls}
\end{equation}

where the hyperparameters $\lambda_{box\_NWD} = 0.05$, $\lambda_{obj} = 1.0$, and $\lambda_{cls} = 0.5$ are empirically set to ensure that the network prioritizes differentiating valid foreground sources from the noise-dominated background. The specifics of each component are detailed below.

\paragraph{Bounding Box Regression Loss ($\mathcal{L}_{box\_NWD}$)}
For small boxes, an offset of one or two pixels can produce a large change in Intersection over Union (IoU), including zero overlap. We use the Normalized Wasserstein Distance \cite[NWD;][]{XU202279} to measure differences in box position and size even for non-overlapping boxes. Each predicted or reference box is represented mathematically by a two-dimensional Gaussian, $\mathcal{N}_p(\mu_p,\Sigma_p)$ or $\mathcal{N}_g(\mu_g,\Sigma_g)$.

For a bounding box $(x_c,y_c, w, h)$, the mean is $\mu = [x_c,y_c]^T$ and the covariance is defined as $\Sigma=diag(w^2/4,h^2/4)$. The second-order Wasserstein distance $W_2^2$ is computed as:

\begin{equation}\label{equ:Wasserstein}
W_2^2(\mathcal{N}_p, \mathcal{N}_g) = ||\mu_p - \mu_g||_2^2 + \text{Tr}\left(\Sigma_p + \Sigma_g - 2(\Sigma_p^{1/2}\Sigma_g\Sigma_p^{1/2})^{1/2}\right)
\end{equation}

For the diagonal covariances above, $W_2^2=\|\mu_p-\mu_g\|_2^2+(w_p-w_g)^2/4+(h_p-h_g)^2/4$. Thus, the distance remains informative when the boxes do not overlap. The normalized regression loss is:

\begin{equation}\label{equ:nwd_loss}
\mathcal{L}_{box\_NWD} = 1 - \exp\left(-\frac{\sqrt{W_2^2(\mathcal{N}_p, \mathcal{N}_g)}}{C_{constant}}\right)
\end{equation}
We adopt the empirical normalization constant $C_{constant}=12.8$ \citep{XU202279}. The distance was computed using bounding-box coordinates in input-image pixels, with $C_{\rm constant}=12.8$ expressed on the same scale.

\paragraph{Objectness Loss ($\mathcal{L}_{obj}$)}
This loss supervises foreground objectness. Its primary role is to separate all potential foreground objects from the empty, noise-dominated \texttt{background}. Crucially, at this stage, the "foreground" encompasses both genuine \texttt{stellar} sources and structural \texttt{fake sources}. To counteract the extreme foreground-background imbalance, we formulate $\mathcal{L}_{obj}$ using a dynamically scaled Focal Loss:

\begin{equation}\label{equ:focal_loss}
\mathcal{L}_{obj} = \sum_{i=2}^{5} w_i \left[ - \frac{1}{N} \sum_{j} \alpha_t (1 - p_t)^\gamma \log(p_t) \right]_i
\end{equation}
where $i \in \{2, 3, 4, 5\}$ denotes the detection head level, and $w_i$ is its assigned weight (with the highest weight $w_2$ assigned to the micro-scale head, giving greater weight to the high-resolution detection head). For each level, $N$ is the total number of anchors and $p_t$ denotes the probability assigned to the target label (for binary targets, $p_t=p$ for positives and $p_t=1-p$ for negatives), $\alpha_t$ balances positive and negative samples, and $\gamma$ is the focusing parameter.

\paragraph{Stellar Classification Loss ($\mathcal{L}_{cls}$)}
The classification term $\mathcal{L}_{cls}$ supervises the class of labeled foreground targets. It is specifically designed to distinguish between the two surviving types of foreground objects: genuine \texttt{stellar} profiles and \texttt{fake source}. We utilize standard Binary Cross-Entropy (BCE):

\begin{equation}\label{equ:cls_loss}
\mathcal{L}_{cls} = - \frac{1}{N_{pos}} \sum_{j=1}^{N_{pos}} \left[ y_j \log(p_j) + (1 - y_j) \log(1 - p_j) \right]
\end{equation}
where $N_{pos}$ is the number of positive foreground samples, $y_j \in \{0, 1\}$ is the ground-truth stellar label (indicating $1$ for \texttt{stellar} and $0$ for \texttt{fake source}), and $p_j$ is the predicted classification probability.

\subsection{Cross-Archive Gaia-guided Fine-tuning}\label{sec:fine-tune}

When directly applying the baseline Plate-YOLO model pre-trained on specific source plates to a new historical archive in a zero-shot manner, detection performance can degrade due to domain shift. Nevertheless, when feeding these initial zero-shot detections into an astrometric solver (e.g., Astrometry.net), some target plates still retain enough valid stellar coordinates to successfully solve their astrometric calibration and obtain a WCS. 

To bridge this cross-domain gap without requiring manual annotation, we introduce a self-calibration strategy that exploits this partial success (the detailed pipeline is discussed in Section \ref{sec:direct_application}). Specifically, we construct an archive-specific dataset using a subset of plates with successful astrometric calibration. On these plates, we first apply a high-sensitivity thresholding algorithm to extract all potential sources. Cross-matching candidates with epoch-propagated Gaia positions projected through the WCS provides catalog-based labels, subject to the matching and catalog-selection limitations in Section~\ref{sec:cutout}. This defines the target dataset $\mathcal{D}_{t}=\{(x_{j}^{t},y_{j}^{t})\}$.

These high-quality labeled samples from the target domain ($\mathcal{D}_{t}$) are then used alongside the source dataset ($\mathcal{D}_{s}$) to fine-tune the Plate-YOLO detector. The regularization term penalizes deviations from the source-model weights, while joint training on $\mathcal{D}_s$ and $\mathcal{D}_t$ is intended to mitigate forgetting during target-archive adaptation.

\begin{equation}
\mathcal{L}(\theta) = \mathcal{L}_{det}(\mathcal{D}_{s}) + \lambda_{t}\mathcal{L}_{det}(\mathcal{D}_{t}) + \lambda_{reg} ||\theta-\theta_{s}||^{2}
\end{equation}

where $\mathcal{L}_{det}$ denotes the total joint detection loss (as defined in Eq. \ref{equ:loss}), $\theta$ represents the current network weights, and $\theta_{s}$ represents the frozen baseline weights of the pre-trained source domain model. The hyperparameters $\lambda_{t}$ and $\lambda_{reg}$ balance the domain adaptation and network regularization, which are empirically set to $\lambda_{t} = 1.0$ and $\lambda_{reg} = 0.01$ in our experiments. The regularization term acts as a structural constraint to prevent catastrophic forgetting; it preserves the foundational source detection capability learned from the original archive, while the target loss component allows the framework to rapidly adapt to the unique optical properties and distinct fake source distributions of the new historical archive.

\subsection{Instrumental Brightness Estimation for Source Ranking}\label{sec:photometric}

The extracted positions are supplied to Astrometry.net for astrometric solving. Relative brightness is used to order the input catalog and prioritize candidate anchors. We estimate an instrumental brightness proxy from the original, unstretched FITS image; it is intended for source ranking rather than calibrated photometry.

The geometric center of a predicted box $[x_{min},y_{min},x_{max},y_{max}]$ is
\begin{equation}
    x_c = \frac{x_{min} + x_{max}}{2}, \qquad
    y_c = \frac{y_{min} + y_{max}}{2}.
\end{equation}

The geometric center of the predicted bounding box was used directly as the aperture center, without centroid refinement or recentering on the original unstretched FITS image. We defined an elliptical measurement aperture from the predicted width $w=x_{max}-x_{min}$ and height $h=y_{max}-y_{min}$, with semi-axes $(a,b)=0.75\times(w,h)$. This empirical scale was applied consistently to all detections, and the resulting image-value sum was used only to rank sources for astrometric solving rather than to estimate their total flux.

The local background $B$ was estimated as the median pixel value within a concentric elliptical annulus with inner semi-axes $(a_{\rm in},b_{\rm in})=1.25\times(w,h)$ and outer semi-axes $(a_{\rm out},b_{\rm out})=1.75\times(w,h)$. Detections with non-positive background-subtracted sums were retained but assigned the lowest ranking.

The background-subtracted image-value sum $F$ is computed on the original, unstretched FITS image:
\begin{equation}
    F = \sum_{\frac{(x-x_c)^2}{a^2} + \frac{(y-y_c)^2}{b^2} \leq 1} 
    \left( I(x,y) - B \right),
\end{equation}

For $F>0$, the instrumental magnitude is defined as $m_{\rm inst}=-2.5\log_{10}(F)+C_{\rm mag}$, where the additive constant affects only the zero point and not the within-plate ordering. The detections were ranked by $m_{\rm inst}$, and only the positions of the 500 highest-ranked sources were passed to \texttt{Astrometry.net} for initial geometric matching. The solver used search depths of $10, 20, \ldots, 100$, a CPU-time limit of 600 s, and a search radius of $1^\circ$ around the supplied pointing coordinates. This quantity therefore serves only as an auxiliary ranking proxy on the scanned-image scale and is not interpreted as a calibrated apparent magnitude.

\section{Application of Plate-YOLO: From Chinese to UBAI plate archives}\label{sec:experiments}

This section shows the application of the Plate-YOLO framework on Chinese and UBAI plate archives. 
We first assess the network's baseline detection capabilities and fake source filtering robustness on the well-characterized Chinese plate archive. Subsequently, we tackle the cross-domain generalization challenge by applying this pre-trained model to the highly heterogeneous UBAI plates, testing both its zero-shot inference and fine-tuned astrometric yields.

\subsection{Plate-YOLO Pre-training and Baseline Evaluation}\label{sec:baseline}

Plate-YOLO was trained on the 2,000-plate Chinese training set for 200 epochs. After each epoch, we evaluated the total detection loss defined in Equation~(\ref{equ:loss}) on the independent 500-plate validation set. We selected the checkpoint with the lowest total validation loss and refer to it hereafter as the \textit{Base\_Model}.

We evaluated the selected \textit{Base\_Model} on the 500-plate Chinese test set. Among stellar-labeled candidates, 88\% are detected and correctly classified, 1\% are classified as non-stellar, and 11\% remain undetected. Among non-stellar-labeled candidates, 7\% are classified as stellar, 26\% as non-stellar, and 67\% remain undetected and are assigned to the background category in this evaluation. This strong background suppression is consistent with the heterogeneous nature of the non-stellar class, which includes artifacts and noise peaks lacking a stable stellar PSF. Together, non-stellar classification and background suppression exclude 93\% (26\%+67\%) of non-stellar-labeled candidates from the stellar catalog, reducing spurious inputs to the astrometric solver.

Astrometric solving benefits from a reliable set of sources distributed across the field. Rejecting artifacts can improve the input catalog, whereas omitting stars can still be harmful if too few anchors remain or their spatial coverage becomes inadequate.

\subsection{Zero-Shot and Fine-Tuning Performance}\label{sec:direct_application}

To evaluate the zero-shot performance of the baseline Plate-YOLO model (\textit{Base\_Model}), which was trained exclusively on the Chinese dataset, we applied it to the 150 uncalibrated UBAI plates.
Despite severe structural differences, relying on the instrumental-brightness ranking from the instrumental magnitude estimation method proposed in Section \ref{sec:photometric}, the model's robust feature extraction enabled Astrometry.net to derive WCS solutions for 40 of these plates successfully. While this provided a necessary foothold in the target domain plates, 110 plates remained unsolved. Differences between archives, image degradation, and solver constraints may all contribute to these failures.

To bridge this domain gap and improve the detection depth limit, we applied the Gaia-guided fine-tuning strategy detailed in Section \ref{sec:fine-tune}. By cross-matching the 40 initially calibrated plates with the Gaia DR3 reference catalogs, we automatically constructed a Gaia-labeled target-domain dataset, $\mathcal{D}_{t} = \{(x_j^t, y_j^t)\}_{j=1}^{N_t}$ (see details in Section \ref{sec:cutout}). The 40 initially solved plates provided 45,800 target-domain training patches.

Building upon the previously established \textit{Base\_Model}, the network was fine-tuned for an additional 50 epochs to yield the \textit{Fine-tuned\_Model}. Optimization was performed using SGD with a momentum of 0.937 (standard YOLOv5 parameter settings). A conservative initial learning rate of $1 \times 10^{-3}$ was adopted, modulated by a cosine annealing schedule with a 3-epoch linear warmup. To mitigate catastrophic forgetting of the generalized representations learned from the source domain, the data loader was configured to sample training batches (batch size of 32) at a strict 1:1 ratio from both the initial Chinese archive ($\mathcal{D}_{s}$) and the new UBAI application scenario ($\mathcal{D}_{t}$). The checkpoint from the final epoch was used as the \textit{Fine-tune\_Model}.

\begin{figure}
\centering
\includegraphics[width=0.75\columnwidth]{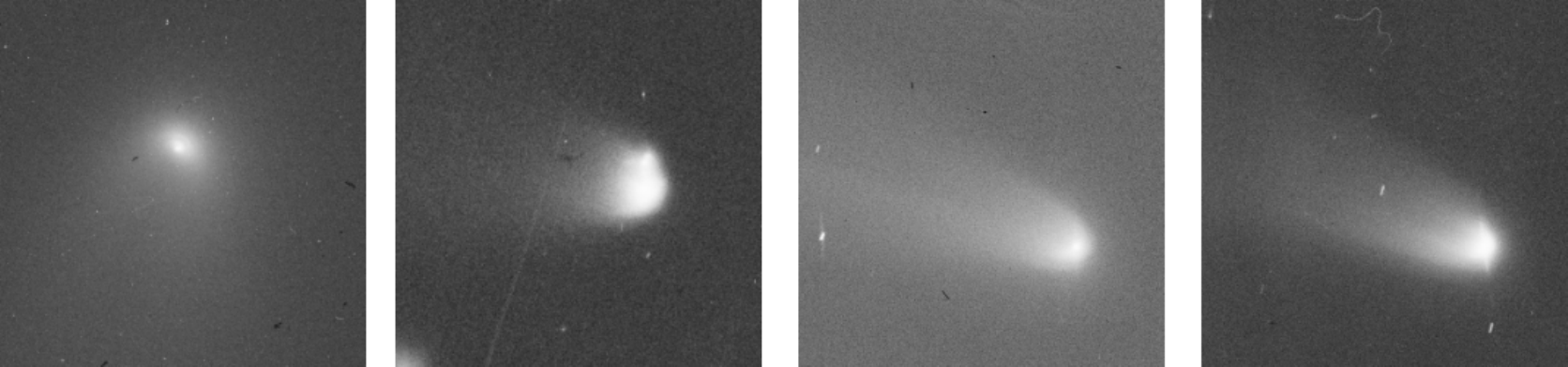}
\caption{The four comet-target plates that remained unsolved after Plate-YOLO fine-tuning.}
\label{fig:failed_calibrations}
\end{figure}

The fine-tuned detector combined with brightness ordering and astrometric solving recovered 106 of the 110 plates. Including the initial 40 solutions, the overall pipeline recovered 146 of the 150 plates (97.3\%). Visual inspection identified the remaining four failures as comet observations: plates x01376, x01397, x01525, and x01527, shown in Figure~\ref{fig:failed_calibrations}. The comet occupies a substantial fraction of the field, while comet tracking produces trails in the background stars, leaving too few usable stellar anchors for reliable astrometric calibration.

\subsection{Science performance of Plate-YOLO detector}

\subsubsection{Detection Completeness and Catalog Purity}

We evaluated detection performance on the 106 newly recovered plates, which were excluded from training and fine-tuning. Using the recovered WCS solutions, we cross-matched the \textsc{SExtractor} and Plate-YOLO catalogs with Gaia DR3 positions propagated to each observation epoch. The Gaia-confirmed \textsc{SExtractor} catalog provides the baseline for relative completeness, while the Gaia-matched fraction of each detection catalog measures its purity. Because rigorous plate photometric calibration is complex \citep{Ma_2025}, instrumental magnitudes serve only as approximate brightness indicators, without correction for plate-to-plate zero-point differences.

\begin{figure*}
\centering

\begin{subfigure}[t]{0.46\textwidth}
    \centering
    \includegraphics[width=\linewidth]{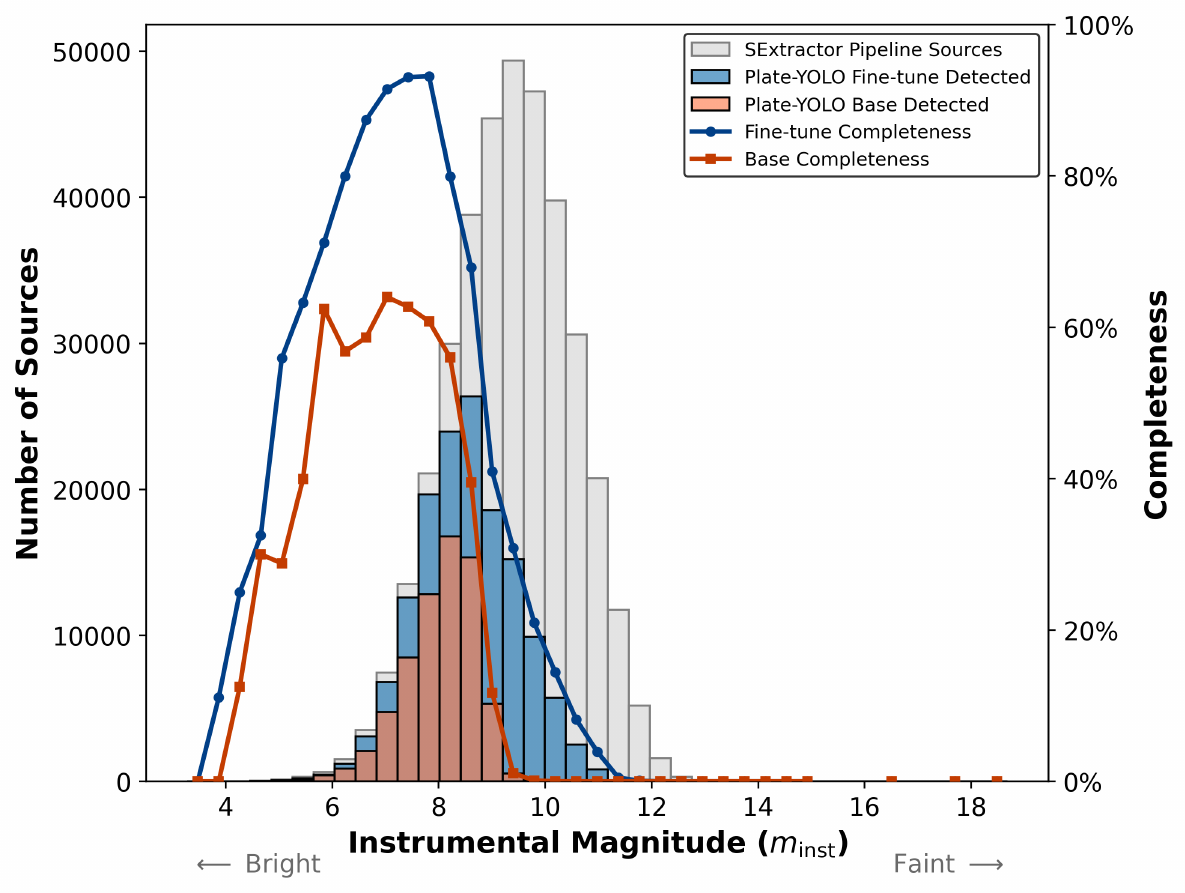}
    \caption{Instrumental-magnitude distributions and detection completeness.}
    \label{fig:magnitude_bias_completeness}
\end{subfigure}
\hfill
\begin{subfigure}[t]{0.5\textwidth}
    \centering
    \includegraphics[width=\linewidth]{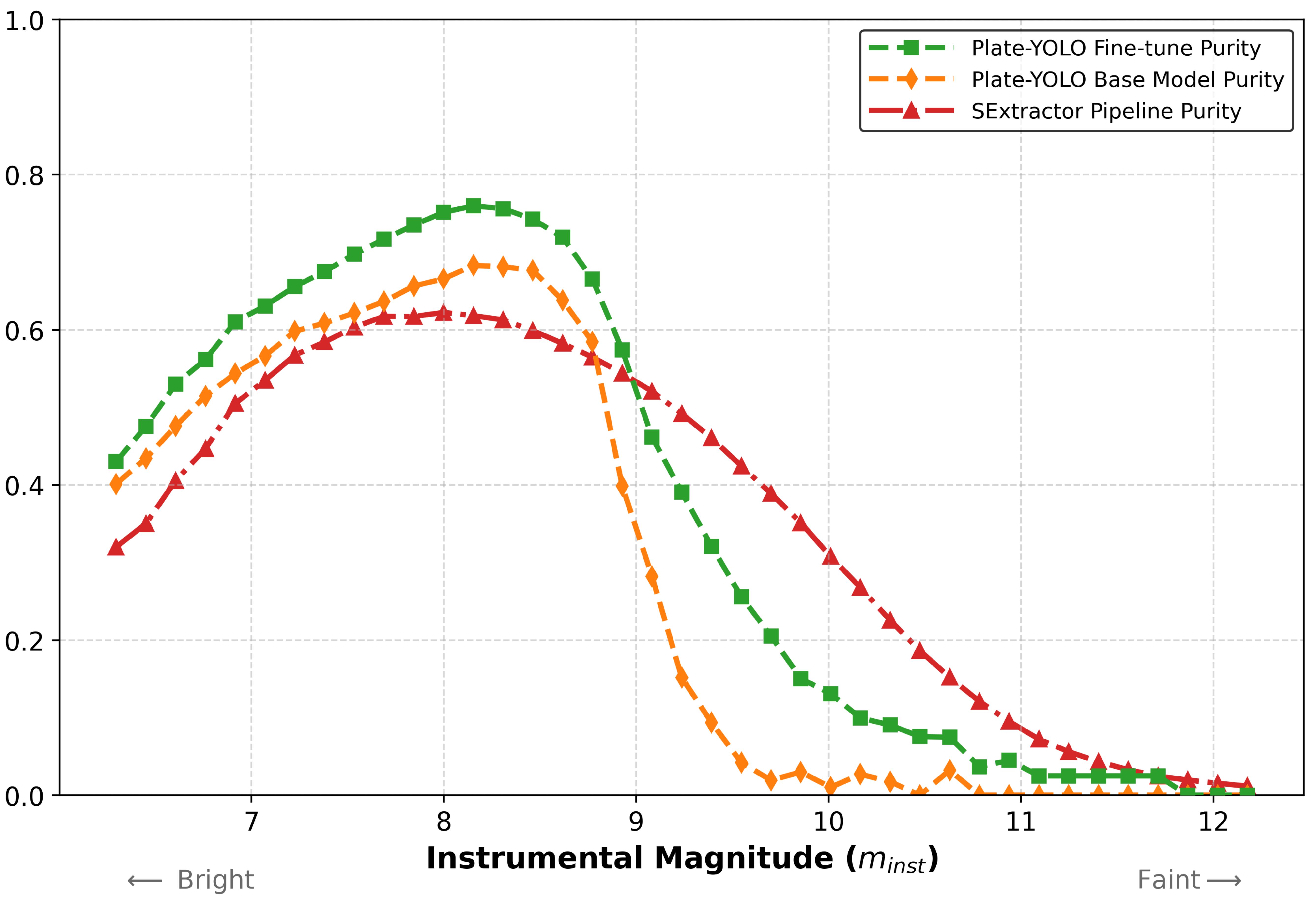}
    \caption{Catalog purity versus instrumental magnitude.}
    \label{fig:magnitude_bias_purity}
\end{subfigure}

\caption{Magnitude-dependent detection performance on the 106 newly recovered plates.
(a) Distributions of Gaia-matched sources detected by \textsc{SExtractor}, the \textit{Base\_Model}, and the \textit{Fine-tuned\_Model}; the curves show the ratio of Gaia-matched Plate-YOLO detections to Gaia-matched \textsc{SExtractor} sources in each magnitude bin.
(b) Purity of the conventional \textsc{SExtractor}-based pipeline, \textit{Base\_Model}, and \textit{Fine-tuned\_Model}.}
\label{fig:magnitude_performance}
\end{figure*}

Figure~\ref{fig:magnitude_bias_completeness} compares the source distributions and magnitude-dependent relative completeness of the two Plate-YOLO models. We adopt a low \textsc{SExtractor} detection threshold of $1.5\sigma$ and treat the Gaia-confirmed detections as an approximately complete reference sample of stars detectable on these plates. In each instrumental-magnitude bin, we calculate the ratio of Gaia-matched Plate-YOLO stellar detections to Gaia-confirmed \textsc{SExtractor} sources. The \textit{Fine-tuned\_Model} reaches a peak ratio of about 95\%, compared with about 65\% for the \textit{Base\_Model}, and stays above 60\% across most of the well-populated bright-source range. Fine-tuning also increases the relative number of faint detections. Beyond $m_{\rm inst}\approx8.5$, the ratio progressively decreases as faint stellar profiles become more difficult to distinguish from background fluctuations.


Figure~\ref{fig:magnitude_bias_purity} compares magnitude-dependent catalog purity. For each Plate-YOLO model, purity is defined as the number of stellar detections matched to Gaia DR3 divided by the total number of detections classified as stellar in the same instrumental-magnitude bin. For the conventional pipeline, purity is calculated analogously as the number of \textsc{SExtractor} detections matched to Gaia DR3 divided by the total number of \textsc{SExtractor} detections in that bin. The \textit{Fine-tuned\_Model} achieves higher purity than the \textit{Base\_Model} and the conventional pipeline over much of the brighter magnitude range. This higher fraction of Gaia-confirmed sources among highly ranked detections may facilitate geometric matching by \texttt{Astrometry.net}.

We further assessed the brightness-ranking consistency of the instrumental-magnitude estimator described on 106 newly recovered plates, selecting up to 500 Gaia-matched sources per plate from each catalog in order of increasing Gaia G magnitude (or all available sources if fewer than 500 were matched). Plate-YOLO achieved a slightly higher median Spearman correlation with G than \textsc{SExtractor}: 0.637 versus 0.603, with interquartile ranges of 0.532–0.765 and 0.537–0.854, respectively. Both catalogs show positive rank correlations with Gaia G, supporting the use of instrumental magnitudes as approximate brightness-ordering proxies. Because the selected source subsets can differ, these results do not establish direct rank agreement between the two methods or superior photometric accuracy.

\subsubsection{Astrometric Quality of the Newly Recovered Plates}

We assessed the astrometric quality of the 106 newly recovered plates using positional residuals relative to Gaia DR3. For matched source $i$, we define
\begin{equation}
\Delta\alpha_i^* =
(\alpha_i-\alpha_{{\rm ref},i})\cos\delta_{{\rm ref},i},
\qquad
\Delta\delta_i = \delta_i-\delta_{{\rm ref},i},
\end{equation}
where $(\alpha_i,\delta_i)$ are the coordinates obtained from the recovered WCS and $(\alpha_{{\rm ref},i},\delta_{{\rm ref},i})$ are the Gaia DR3 coordinates propagated to the observation epoch. The factor $\cos\delta_{\mathrm{ref},i}$ converts the right-ascension difference to its local tangent-plane angular separation. Both residual components are expressed in arcseconds. The positional residual is $r_i=[(\Delta\alpha_i^*)^2+(\Delta\delta_i)^2]^{1/2}$, and the plate-level two-dimensional RMS is
\begin{equation}\label{equ:plate_rms}
\mathrm{RMS} = \sqrt{\frac{1}{N}\sum_{i=1}^{N}r_i^2},
\end{equation}
where $N$ is the number of matched sources used in the evaluation.

\begin{figure}
\centering
\includegraphics[width=0.6\columnwidth]{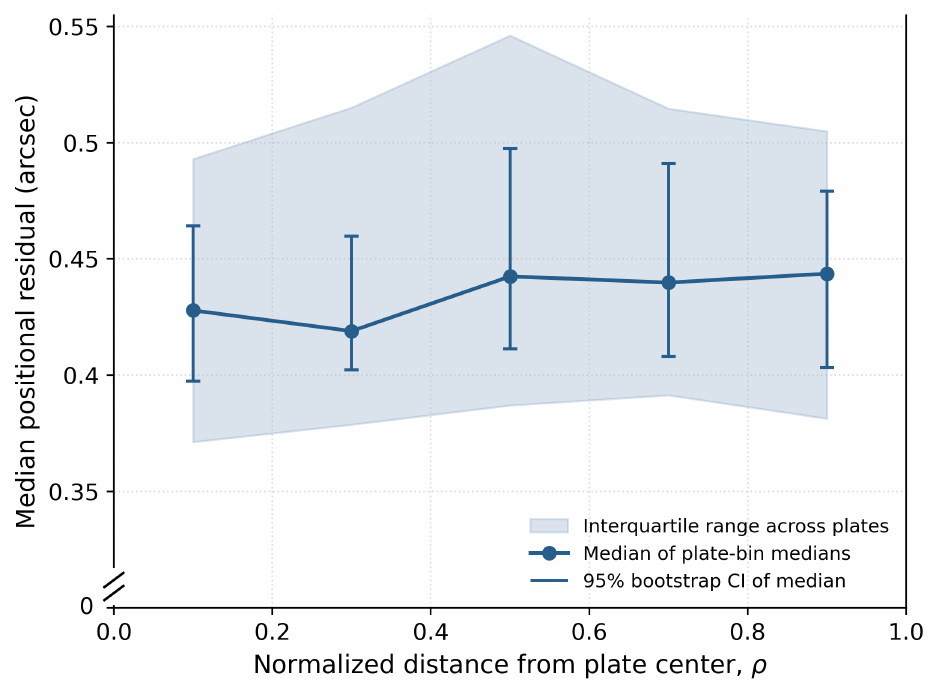} 
\caption{Radial dependence of Gaia DR3 positional residuals for the 106 recovered plates. Points denote equally weighted medians of plate-level bin medians; shading and error bars indicate interquartile ranges and bootstrap 95\% confidence intervals, respectively. The vertical axis is broken below $0.3''$.}
\label{fig:radial_residual_stability}
\end{figure}

Across these plates, the median RMS is $0.563''$ (interquartile range: $0.496''$--$0.635''$), based on a median of 1724 Gaia matches per plate (interquartile range: 602--4253). These residuals are consistent with the stronger image degradation in this challenging sample, including PSF distortions and emulsion defects.

Figure~\ref{fig:radial_residual_stability} examines the radial dependence of the positional residuals. Here, $\rho=\sqrt{(x-x_c)^2+(y-y_c)^2}/[\sqrt{W^2+H^2}/2]$, where $(x_c,y_c)$ is the plate center and $W$ and $H$ are its dimensions. We divided the normalized distance from the plate center, $\rho$, into five bins of width 0.2 and calculated the median residual for each plate and bin containing at least five matched sources. The plotted values are the medians of these plate-level measurements, with equal weight assigned to each contributing plate. The median residual remains approximately $0.42''$--$0.44''$ across the radial bins, showing no pronounced outward increase.

\section{Discussion}\label{sec:discussion}

To compare processing time and astrometric residuals, we randomly selected 500 previously solved UBAI plates from the 2,330 plates analyzed by \citet{10.1093/mnras/stag335} and reprocessed them with Plate-YOLO. These approximately $8000\times8000$-pixel images form the benchmark used in the following comparisons.

\subsection{Processing Speed}\label{sec:processing_speed}

Table~\ref{tab:speed_compare} summarizes the mean processing times for the 500-plate benchmark. Plate-YOLO required 2.86~s for image preprocessing, 3.75~s for inference, and 6.61~s for detection per plate. This stage performs both source detection and stellar classification, whereas \textsc{SExtractor} required 16.73~s to detect candidate sources. Plate-YOLO completes this broader task 2.53 times faster.

\begin{table}[h]
	\centering
	\caption{Comparison of average computational time per plate between the SExtractor and the Plate-YOLO framework in \textbf{500} UBAI plates.}
	\label{tab:speed_compare}
    \setlength{\tabcolsep}{1.2mm}{
	\begin{tabular}{c|c|ccc|c|c}
		\hline
        \multirow{2}{*}{Method} & \multirow{2}{*}{Hardware} & \multicolumn{3}{c|}{Source extraction} & \multirow{2}{*}{Calibration} & \multirow{2}{*}{Total} \\
        \cline{3-5}
         & & Pre pro & Detection & Subtotal & & \\
        \hline
        Plate-YOLO & RTX 3090(GPU) & 2.86 & 3.75 & 6.61 & 0.79 & 7.40\\
        SExtractor & i9-10900(CPU) & - & - & 16.73 & 2.63 & 19.36\\
        \hline
	\end{tabular}}
\end{table}

Astrometry.net solving time decreased from 2.63 to 0.79~s with the Plate-YOLO catalogs. The measured extraction and solving stages required 19.36~s for the conventional pipeline and 7.40~s for Plate-YOLO, corresponding to a 2.62-fold acceleration. Because \textsc{SExtractor} ran on an Intel i9-10900 CPU and Plate-YOLO on an NVIDIA RTX 3090 GPU, these values represent the practical performance of the adopted configurations rather than a hardware-independent comparison.

\subsection{Astrometric Residuals}\label{sec:astrometric_residuals}

We compared the two pipelines on the same 500-plate benchmark using the plate-level RMS defined in Equation~\ref{equ:plate_rms}. Both pipelines used the same Gaia DR3 reference selection, matching criteria, clipping procedure, and WCS model. Each RMS was calculated from the corresponding pipeline's own Gaia-matched sources, so the paired comparison evaluates the resulting solutions, including differences in source selection.

\begin{figure}
    \centering
    \includegraphics[width=0.6\columnwidth]{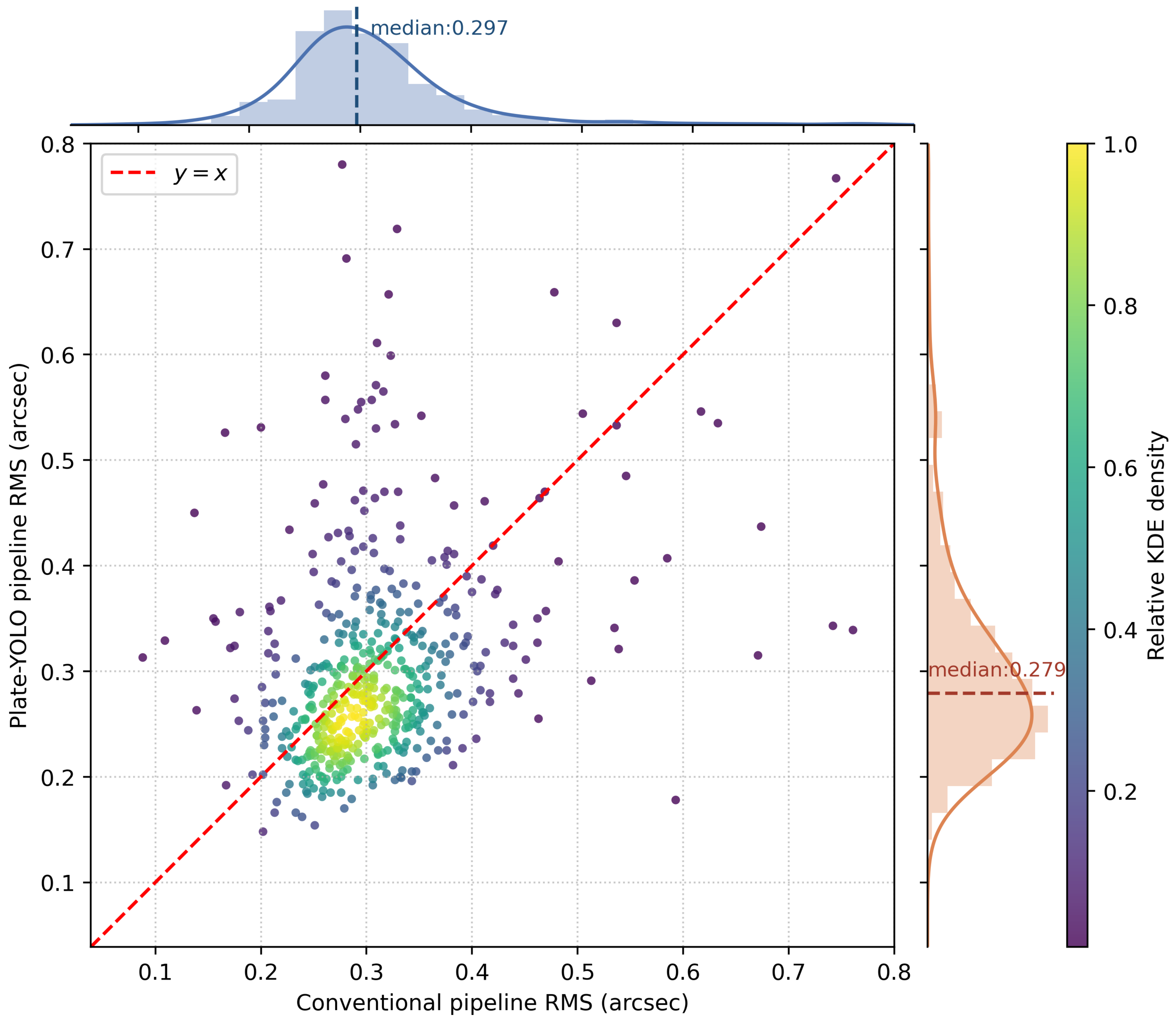}
    \caption{Comparison of astrometric calibration RMS between Plate-YOLO object detection and conventional extraction methods in 500 previously solved UBAI plates. The $y=x$ line indicates equality, and colors represent the Gaussian kernel density estimate normalized to its maximum.}
    \label{fig:rms_comparison}
\end{figure}

Figure~\ref{fig:rms_comparison} compares the paired RMS measurements for the 500 previously solved plates, with the marginal distributions showing the median RMS of each pipeline. Plate-YOLO achieved a smaller RMS for 308 plates (61.6\%). Defining $\Delta_i={\rm RMS}_{{\rm Plate\text{-}YOLO},i}-{\rm RMS}_{{\rm conventional},i}$, the median paired difference was $-0.022$~arcsec, with an interquartile range of $[-0.064'',\,0.033'']$~arcsec and a bootstrap 95\% confidence interval of $[-0.030'',\,-0.013'']$~arcsec. These results indicate a modest overall reduction in the reported plate-level residuals for the Plate-YOLO pipeline, although the improvement is not uniform across plates and does not isolate WCS accuracy from source-selection effects.

\section{Conclusions}\label{sec:conclusion}

Plate-YOLO integrates source detection and stellar classification for degraded historical photographic plates using SPD feature rearrangement, SimAM attention, a P2 detection head, and NWD regression. The resulting stellar catalogs are ranked by an instrumental-brightness proxy and passed directly to the astrometric solver.

Among 150 UBAI plates unsolved by the conventional pipeline, the Chinese-trained model recovered 40 solutions and Gaia-guided fine-tuning recovered a further 106, giving an overall success rate of 146/150 (97.3\%). The newly recovered solutions have a median plate-level RMS of $0.563''$, with no pronounced outward increase in the median positional residuals. On the tested hardware, Plate-YOLO achieved 2.53-fold faster source extraction and a 2.62-fold speedup for the combined extraction and astrometric-solving workflow.

These results demonstrate that archive-specific adaptation can recover astrometric information from plates affected by optical distortions, emulsion defects, stellar trailing, and other forms of degradation that hinder conventional source extraction. Plate-YOLO therefore provides a practical route toward large-scale reprocessing of heterogeneous historical plate archives while reducing the need for archive-specific manual parameter tuning. Its present performance remains dependent on the available reference catalogs, training-label construction, and the plate populations examined here, motivating further validation across additional instruments and archives before broader deployment.

\begin{acknowledgments}
This work is supported by the National Key R\&D Program of China (Grant No. 2025YFF0510603), the National Natural Science Foundation of China (Grant No. 12473070) and the International Partnership Program of the Chinese Academy of Sciences (Grant No. 018GJHZ2023110GC).
\end{acknowledgments}

\appendix

\section{Ablation Study of Plate-YOLO}\label{sec:ablation}

To systematically validate the structural modifications introduced in Plate-YOLO, we conducted an ablation study using only the Chinese historical-plate cutouts described in Section~\ref{sec:cutout}. The training, validation, and test sets are from this dataset, thereby excluding cross-domain effects from the architectural comparison. Using a standard YOLOv5 baseline, we incrementally integrated our proposed components and evaluated them on an independent test set to quantify their incremental effects to raw detection capability. The performance gains are detailed in Table~\ref{tab:ablation_arch}.

\begin{table}[h]
\centering
\caption{Ablation study of the architectural components on the test dataset of the Chinese plate.}\label{tab:ablation_arch}
\setlength{\tabcolsep}{0.4mm}{
\begin{tabular}{llccc}
\toprule
Model & Added & \multicolumn{3}{c}{Stellar}  \\
\cline{3-5}
Variant & Component & Precision (\%) & Recall (\%) & AP@0.5 \\
\midrule
Baseline & None & 85.6 & 78.4 & 0.754 \\
Variant 1 & + NWD Loss & 91.8 & 82.3 & 0.803 \\
Variant 2 & + SPD Module & 92.4 & 85.1 & 0.826 \\
Variant 3 & + SimAM & 95.5 & 84.8 & 0.841 \\
Plate-YOLO & + P2 Head & \textbf{97.0} & \textbf{88.0} & \textbf{0.874} \\
\bottomrule
\end{tabular}}
\tablecomments{Baseline denotes the unmodified YOLOv5 model, and components are added cumulatively. Precision, recall, and AP@0.5 are reported for the stellar class on the Chinese test patches, with AP evaluated at an IoU threshold of 0.5. We focus on stellar detection because the resulting stellar catalog is input to the astrometric solver.}
\end{table}

The unmodified YOLOv5 baseline gives the reported precision of 85.6\% and recall of 78.4\%. The following results describe the cumulative experiment; they do not isolate the contribution of each component from interactions with previously added components.

\begin{itemize}

\item \textbf{NWD Loss:}
NWD compares bounding-box centers and sizes without requiring spatial overlap, making the regression loss less sensitive to small pixel-level offsets for tiny targets. Adding NWD increases the reported precision to 91.8\% and recall to 82.3\%.

\item \textbf{Space-to-Depth (SPD) Module:}
The SPD module rearranges spatial samples into the channel dimension before subsequent feature extraction, reducing information loss associated with direct downsampling. Its inclusion increases the reported recall to 85.1\%.

\item \textbf{SimAM Attention:}
Adding SimAM increases the reported precision to 95.5\%, with a recall of 84.8\% and an AP@0.5 of 0.841.

\item \textbf{Micro-Scale Head (P2):}
The P2 head introduces a finer-resolution detection scale for small sources. Its inclusion increases the reported precision to 97.0\%, recall to 88.0\%, and AP@0.5 to 0.874. The improvement is consistent with better localization of small targets.

\end{itemize}

Overall, AP@0.5 increases progressively as the components are added. All variants use the same data partitions and training configuration, so the experiment measures the incremental effects of the cumulative modifications under a fixed experimental setting.

\bibliography{ajref}{}
\bibliographystyle{aasjournalv7}



\end{document}